\documentclass{elsart}

\usepackage{natbib}

 \usepackage{graphicx}

\usepackage{amssymb}

\begin{document}

\begin{frontmatter}



\title{Silicon sensors implemented on p-type substrates for high radiation resistance application }


\author{Marina Artuso}

\address{Syracuse University, Syracuse, NY 13244, USA}

\begin{abstract}
Silicon based micropattern detectors are
 essential elements of
modern high energy physics experiments. Cost effectiveness and high
radiation resistance are two important requirements for technologies
to be used in inner tracking devices. Processes based on p-type
substrates have very strong appeal for these applications. Recent
results and prototype efforts under way are reviewed.

\end{abstract}


\end{frontmatter}

\section{Introduction}
The CERN Large Hadron Collider (LHC) is poised to start operation
soon and ramp up to a luminosity of 10$^{34}$ cm$^{-2}$s$^{-1}$. To
be able to cope with the high levels of radiation expected in the
inner tracking devices a new technology has been developed: the so
called ``n-on-n'' silicon micropattern detectors have replaced the
conventional ``p-on-n'' used for decades in Si microstrip devices.
In this approach, the charge signal collecting electrodes are
implemented with $n^+$ implants on an $n$-type substrate, while the
$p^+$ implant extends throughout the back plane of the device.  One
of the reasons underlying this choice is the discovery of
``type-inversion'' \cite{type-inversion}. Radiation damage
introduces electrically active defects in the band gap that change
several sensor properties , in particular they alter the effective
doping concentration $N_{eff}$. The study of the $N_{eff}$ evolution
lead to the discovery of the phenomenon of ``charge inversion'',
naively defined as a change in the $N_{eff}$ sign (transformation
into a ``p-type'' equivalent substrate). This effective doping
concentration is described with the equation
\begin{equation}
N_{eff}=N_{eff}(0)e^{-c\Phi}-\beta\Phi, \end{equation}

where $N(0)$ is the initial doping concentration, $c$ and $\beta$
are empirical parameters accounting for the donor removal and the
acceptor-like defect addition rates as a function of the radiation
dose $\Phi$. This is an effective parametrization;  studies have
shown that a ``double-junction'' model
\cite{casse-type-inv,li-type-inv,beattie-type-inv}, where the main
junction is at the $n^+$-bulk interface, provides a better
description of the device properties after high radiation doses.
After type inversion, ``n-on-n'' device achieve much better charge
collection efficiency that conventional devices. Thus Atlas
\cite{atlas},  CMS \cite{cms}, and LHCb\cite{lhcb} use this
technology for their tracking subsystems exposed to the largest
radiation doses. An upgrade of LHC (sLHC) is planned to achieve a
luminosity of the order of 10$^{35}$ cm$^2$s$^{-1}$, a factor of 10
higher than the LHC design luminosity. A corresponding increase in
the expected radiation doses is a natural consequence. Thus new
technologies are being studied to be used in the hottest detector
regions.

One of the drawbacks of the ``n-on-n'' technology is the high field
at the backplane junction in the early stages of operation of the
detectors, prior to radiation damage. Because of this, ``n-on-n''
detectors are produced with a double-sided technology: implants in
the back side include complex guard ring structures that have the
goal of providing a controlled drop of the voltage from the biasing
implant to the edge of the device. Moreover, the high field region
produces a quickest charge collection and thus it is desirable to
have it always close to the signal collection electrode. Although
many devices have been produced with the ``n-on-n'' technology,
options that could overcome these limitations have been sought. High
resistivity, detector grade p-type substrates provide a very
promising solution to satisfy the radiation resistance requirements
in a very cost effective manner.

\section{p-type technology}
One of the main technological challenges in the fabrication of
``n-in-p'' devices is the achievement of a good inter-electrode
isolation. This is because the positive charge in the SiO$_2$ oxide
induces an electron accumulation layer at the oxide-Si interface
that would electrically connect all the sensing elements if no
isolation mechanism were introduced. This is an old problem,
addressed for the first time when double-sided Si microstrip devices
were introduced \cite{batignani}, but it is still challenging the
ingenuity of device designers and foundries. The charge density in
the accumulation layer increases with the radiation dose, up to a
level known as the ``oxide saturation charge''. Interstrip isolation
must guarantee a high inter-electrode resistance throughout the
sensor lifetime, without adversely affecting the electrical
performance of the device. Three approaches commonly used are known
as p-stop, p-spray, and moderated p-spray. The p-stop method is
based on  p-type implants surrounding the n$^+$ collection
electrode. ``n-in-p'' sensors with p-stop interstrip isolation have
been shown to feature higher radiation resistance than sensors of
equal geometry implemented with ``n-in-n'' technology
\cite{lozano-ieee05}. However p-stop implants require an additional
mask, and they suffer from pre-breakdown micro-discharges that
deteriorate the noise performance of the sensors implemented with
this interstrip insulation. A commonly used alternative is the
``p-spray'' technique, where a uniform p-type blanket is implanted
throughout the active surface of the device. The n$^+$ implants have
sufficient dose to overcompensate this p-type layer in the regions
where charge collection electrodes are planned. Thus the additional
mask is no longer needed. On the other hand, in this case it is
necessary to ensure that early breakdowns do not occur, and that an
acceptable inter-electrode insulation is maintained throughout the
detector lifetime. Thus a careful tuning of the p$^+$ implant is
necessary. A hybrid solution that aims at bridging the advantages of
the two methods, the so-called ``moderated p-spray'' is sometimes
used. This solution requires an additional mask, but may ease the
conflicting constraints of avoiding early breakdowns and maintaining
high interstrip resistance at all the radiation levels. Note that
due to the lower hole mobility, the depletion voltage  is three
times higher for p-type substrates than n-type substrates of a given
resistivity.

The high energy physics community has decades of experience with
detector grade n-type substrates, while the p-type option is
relatively new. However, several groups  are studying this
technology. Both the CERN RD50 collaboration\cite{rd50} and the INFN
funded SMARTS collaboration \cite{smarts} have devoted considerable
resources towards a comprehensive study of the properties of a
variety of sensors implemented with this technology.  Sensors have
been produced at research laboratories such as CNM-IMB \cite{cnm},
or ITC-IRST \cite{itc}, as well as at industrial foundries such as
Micron \cite{micron}, and Hamamatsu \cite{hamamatsu}. Thus  a
thorough understanding of the properties of these devices at
different level of radiation is emerging.

\section{Fabrication Details}
  Both
diffusion oxygenated float zone (DOFZ), and magnetic Czochralski
(MCz) Si wafers have been used, with a wide resistivity range and
wafer thicknesses typically between 200 $\mu$m and 300 $\mu$m.
Different isolation techniques have been used. The vast majority of
the efforts has been concentrated on blanket p-spray, but different
p-stop topologies, including field plate modifications \cite{unno},
and moderated p-spray have been implemented.

Earlier wafers contained a large number of test structures, such as
multi-guard ring diodes, MOS capacitors, and gated diodes, and a
smaller number of strip or pixel detectors. Now the effort is
progressing towards implementation of larger scale devices. For
example, the RD50 collaboration has a set of devices being
manufactured at Micron on 6 inch wafers that include strip and pixel
detectors suitable for LHCb, ATLAS, and CMS upgrades. In addition,
full size VELO sensors on high resistivity p-type substrates have
been produced, that have been assembled into fully instrumented
modules by the University of Liverpool group.

\section{Performance before and after irradiation}

There is a vast array of studies on the performance before and after
irradiation of the detectors implemented on p-type substrates. Some
examples illustrating the achievements and challenges in this
process will be discussed.

 Blanket p-spray isolation with
lower implant dose \cite{pellegrini:nim566} has been shown to fail
to provide adequate interstrip resistance at a dose of about 50
MRad, which is the total dose expected in the middle region of the
future upgrade of the Atlas detector at Super-LHC. A KEK group
\cite{unno-trento} reported similar findings for lower dose p-spray
and p-stop implants. This group is investigating alternative
isolation techniques to achieve the optimum interstrip resistance,
including the use of field plates over the blocking implant.

\begin{figure}
\begin{center}
\includegraphics*[width=10cm]{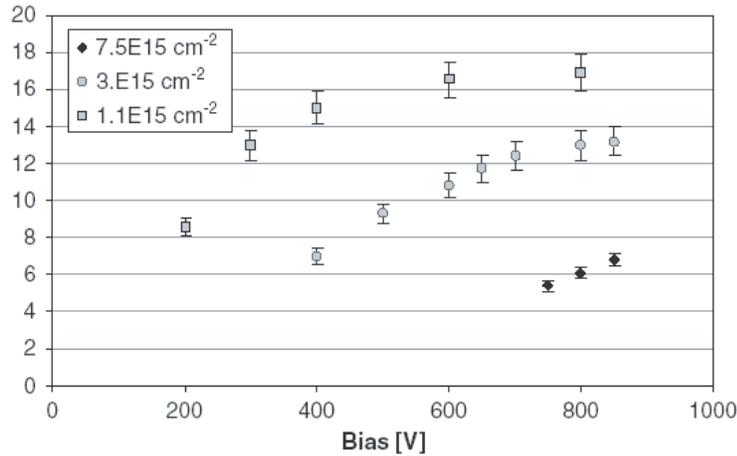}
\end{center}
\vskip -4cm \caption{Charge collection efficiency
versus applied voltage, normalized to pre-irradiation value, of
``n-in-p'' strip detectors. The detector irradiated at $3\times
10^{15}$ pcm$^{-2}$ is a standard p-type substrate, while the others
are oxygen enriched. The measurements \cite{gianluigi} have been
performed at a temperature of -20/25$^\circ$C.\label{lhcb:cc}}
\end{figure}

Other operational aspects that need to be studied as a function of
the radiation dose are the current versus voltage characteristic and
the charge collection efficiency. It has been argued effectively
\cite{gianluigi} that the latter parameter is a much better
predictor of the longevity of any given detector technology than the
full depletion voltage $V_{fd}$ extracted from a capacitance versus
voltage measurement.  While $V_{fd}$ may be not practically
achievable because it would induce thermal runaway in  radiation
damaged sensors,  adequate charge collection efficiency at lower
voltages may be still be achieved. Figure~\ref{lhcb:cc} shows an
example of the charge collection efficiency obtained in a 1x1 cm$^2$
280 $\mu$m thick microstrip detector produced by CNM-IMB using masks
designed at the University of Liverpool. The charge collection
measurement has been performed using a $^{106}$Ru source, that has
an energy deposition comparable to a minimum ionizing particles.
Thus the study provides an absolute charge collection efficiency. As
the noise has been shown not to depend upon the irradiation level,
the signal to noise ratio scales with the charge signal, and,
although it is deteriorating as a function of the radiation dose,
may still be acceptable at the highest levels of radiation
considered.

\begin{figure}
\begin{center}
\includegraphics*[width=12cm]{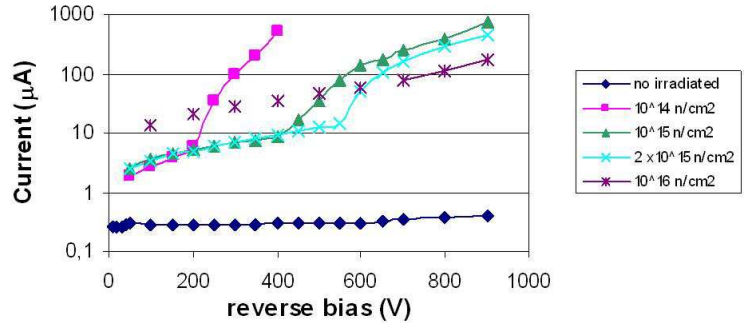}
\end{center}
\vskip -6cm \caption{ Current-Voltage characteristics of ``n-in-p''
microstrip detectors irradiated with neutrons. The
measurements\cite{minano} have been performed at a temperature of
-30$^\circ$C. \label{fig:iv}}
\end{figure}

Other radiation studies have been reported
\cite{hara,liverpool,segneri}. One of the most recent is based on
the measurements by a group at IFIC, Valencia, \cite{minano}
  of 4 microstrip detectors
 manufactured by CNM-IMB and
irradiated with neutrons at the TRIGA Mark II reactor in Ljubljana
to different fluences ranging from 10$^{14}$ to 10$^{16}$ n/cm$^2$.
Figure~\ref{fig:iv} shows the current versus voltage characteristic
of irradiated sensors maintained at a temperature of -30$^\circ$ C;
a non irradiated sensor is included for reference. Earlier
micro-discharges appear at lower irradiation doses. This group
investigated the charge collection properties by measuring the
charged signal induced by a pulsed infrared laser (1060 nm). At the
higher fluences the charge collection efficiency does not reach the
plateau corresponding to full depletion, and microdischarges onset
was observed.

First examples of the feasibility of these detectors in real scale
devices have been accomplished. For example, 6x6 cm$^2$ silicon
strip devices designed with ATLAS strip detector geometry
implemented with n$^+$ strips implanted on different substrates have
been constructed and tested \cite{gianluigi}. The LHCb-VELO group
has produced full scale sensors  on 300 $\mu$m thickness high
resistivity p-type substrates: the depletion voltage is of the order
of 100 V, while breakdown is observed at a voltage exceeding 200 V.
A fully instrumented VELO module was tested in the laboratory both
at the university of Liverpool and CERN, and was used in a recent
test beam run including 10 VELO modules. Another large scale
production of large microstrip and pixel modules on 6" wafers is
under way at Micron, with masks developed by the RD50 collaboration.
The VELO beam test data and laboratory and test beam
characterization of the RD50 devices, before and after irradiation,
will further our understanding of the operational and system
properties of detectors built with this technology.

\section{Conclusions}
The examples shown in this paper give some snapshots of the vibrant
R\&D effort to optimize processes based on p-type substrates to
implement micropattern detectors. The ongoing research is making
significant strides towards the development of a cost effective
technology suitable for the highest radiation doses expected at
sLHC.

\section{Acknowledgement}
I would like to thank the organizers for a very enjoyable and
productive conference in a beautiful setting. I would also like to
acknowledge interesting discussions and scientific input from G.
Casse, and M. Lozano. This work was supported by the United States
National Science Foundation.




\end{document}